\RequirePackage{amsthm}
\documentclass[sn-mathphys-num]{sn-jnl}

\usepackage{graphicx}%
\usepackage{multirow}%
\usepackage{amsmath,amssymb,amsfonts}%
\usepackage{amsthm}%
\usepackage{mathrsfs}%
\usepackage[title]{appendix}%
\usepackage{xcolor}%
\usepackage{textcomp}%
\usepackage{manyfoot}%
\usepackage{booktabs}%
\usepackage{algorithm}%
\usepackage{algorithmicx}%
\usepackage{algpseudocode}%
\usepackage{listings}%
\usepackage{pdfpages}%
\usepackage{textcomp}
\usepackage{subfig} 

\theoremstyle{thmstyleone}%
\theoremstyle{thmstyletwo}%

\theoremstyle{thmstylethree}%

\graphicspath{ {./Figures/} }

\begin{document}

\title[Article Title]{High-Magnitude Earthquake Identification Using an Anomaly Detection Approach on HR-GNSS Data}


\author[1,2]{\fnm{Javier} \sur{Quintero-Arenas}}\email{jaqarenas@fias.uni-frankfurt.de}

\author[1]{\fnm{Claudia} \sur{Quinteros-Cartaya}}\email{quinteros@fias.uni-frankfurt.de}

\author[3]{\fnm{Andrea} \sur{Padilla-Lafarga}}\email{andreapadilla@uas.edu.mx}

\author[3]{\fnm{Carlos} \sur{Moraila}}\email{cmoraila@uas.edu.mx}

\author[1,4]{\fnm{Johannes} \sur{Faber}}\email{faber@fias.uni-frankfurt.de}


\author[1,5]{\fnm{Jonas} \sur{Köhler}}\email{jkoehler@fias.uni-frankfurt.de}

\author*[1,5]{\fnm{Nishtha} \sur{Srivastava}}\email{srivastava@fias.uni-frankfurt.de}

\affil[1]{\orgname{Frankfurt Institute for Advanced Studies}, \orgaddress{\street{Ruth-Moufang-Straße 1}, \city{Frankfurt am Main}, \postcode{60438}, \state{Hessen}, \country{Germany}}}

\affil[2]{\orgdiv{Institute of Computer Sciences}, \orgname{Goethe University}, \orgaddress{\street{Robert-Mayer-Str. 11-15}, \city{Frankfurt am Main}, \postcode{60325}, \state{Hessen}, \country{Germany}}}

\affil[3]{\orgdiv{Faculty of the Earth and Space Sciences}, \orgname{Autonomous University of Sinaloa}, \orgaddress{\street{Universitarios Ote. S/N, Cd Universitaria}, \city{Culiacan Rosales}, \postcode{80040}, \state{State}, \country{Mexico}}}

\affil[4]{\orgdiv{Institute for Theoretical Physics}, \orgname{Goethe University}, \orgaddress{\street{Max-von-Laue-Str. 1}, \city{Frankfurt am Main}, \postcode{60438}, \state{Hessen}, \country{Germany}}}

\affil[5]{\orgdiv{Institute of Geosciences}, \orgname{Goethe University}, \orgaddress{\street{Altenhöferallee 1}, \city{Frankfurt am Main}, \postcode{60438}, \state{Hessen}, \country{Germany}}}


\abstract{Earthquake early warning systems are crucial for protecting areas that are subject to these natural disasters. An essential part of these systems is the detection procedure. Traditionally these systems work with seismograph data, but high-rate GNSS data has become a promising alternative for the usage in large earthquake early warning systems. Besides traditional methods, deep learning approaches have gained recent popularity in this field, as they are able to leverage the large amounts of real and synthetic seismic data. Nevertheless, the usage of deep learning on GNSS data remains a comparatively new topic. This work contributes to the field of early warning systems by proposing an autoencoder based deep learning pipeline that aims to be lightweight and customizable for the detection of anomalies \textit{viz.} high magnitude earthquakes in GNSS data. This model, DetEQ, is trained using the noise data recordings from nine stations located in Chile. The detection pipeline encompasses: (i)the generation of an anomaly score using the ground truth and reconstructed output from the autoencoder, (ii) the detection of relevant seismic events through an appropriate threshold, and (iii) the filtering of local events, that would lead to false positives. Robustness of the model was tested on the HR-GNSS real data of 2011 Mw 6.8 Concepci\'on earthquake recorded at six stations. The results highlight the potential of GNSS-based deep learning models for effective earthquake detection.}




\maketitle

\section{Introduction}\label{sec1}

Over recent years, Global Navigation Satellite System (GNSS) data has gained increasing recognition in seismology for its ability to capture the ground displacements directly caused by large earthquakes, which is crucial for both earthquake source characterization and early warning systems \cite{bib_detection},\cite{bib_rupture},\cite{bib11},\cite{bib14}.

Simultaneously, advances in machine learning, particularly deep learning (DL), have been leveraged to explore the efficiency of new algorithms intending to improve earthquake detection systems \cite{bib22}. Several studies have demonstrated that DL algorithms applied to seismic data outperform traditional methods in terms of both accuracy and speed (e.g. \cite{bib_DL_EEW},\cite{bib_creime_rt},\cite{bib_DLSeis}). However, most approaches have concentrated on seismic waveform data recorded from inertial instruments, with limited exploration of DL models using geodetic data for earthquake monitoring. This gap offers an opportunity to investigate GNSS data analysis as a complementary method for earthquake detection.

Processing GNSS data for real-time earthquake detection remains a significant challenge, primarily due to noise and the complexity of real-time data processing \cite{bib_noise},\cite{bib_real_time_1},\cite{bib_GSeisRT}. Despite these difficulties, GNSS data can provide information about permanent ground displacements that complement seismic waveform data, especially in scenarios where traditional seismic networks face limitations, such as insufficient good quality or a lack of near-source seismic data.

In this paper, we propose a DL model, DetEq, designed to detect large earthquakes (Mw $>$ 6) using high-rate GNSS data, with a 1 Hz interval rate (HR-GNSS). The model leverages the geodetic observations in each station in the absence of an earthquake to detect the onset of ground displacements due to significant seismic event which is treated as an anomaly.

The significance of this work lies in its potential to enhance the speed and accuracy of earthquake detection using HR-GNSS and it is expected to facilitate the further integration of DL models for GNSS data analysis into seismic monitoring networks and benefit the management of risk mitigation.

Finally, the proposed model for earthquake detection in this paper, together with an additional DL model for earthquake magnitude estimation \cite{bib_gnss_mag1},\cite{bib_gnss_egu}, will be incorporated into the SAIPy package \cite{bib_saipy}, an open-source Python framework designed for earthquake analysis. This integration will provide a comprehensive toolset for both earthquake detection and magnitude assessment using GNSS data.

\section{Architecture}\label{sec2}
\begin{figure}[h!]
\centering
\includegraphics[scale=0.5]{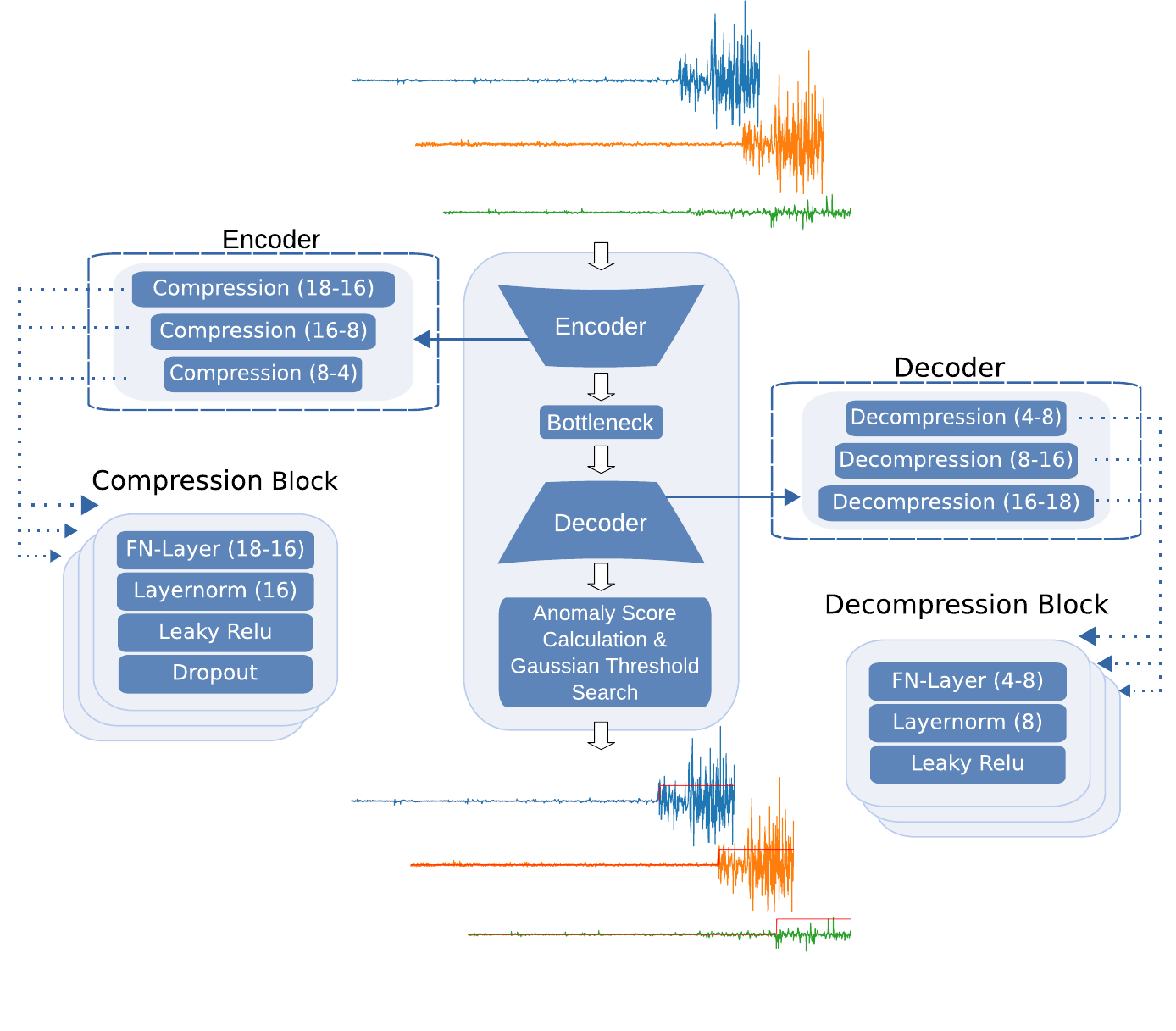}
\caption{The complete detection pipeline takes as input a flattened time series array of 6 seconds and 3 channels resulting in an array of shape (18). The pipeline consists of an autoencoder with encoder and decoder. The subsequent anomaly calculation uses pointwise the $L_2$ norm. Finally the threshold search module specifies the anomaly threshold.}
\label{fig:det_arch}
\end{figure}
The present study utilizes an autoencoder comprising three main components: an encoder block, a bottleneck and a decoder block. The input data for the model has the shape (x, 3) with x being the length of the recording. The second dimension encompasses the three GNSS channels. For processing, the data is cut into six second strips with the shape of (6, 3) and fed to the encoder block, which consists of fully connected layers, normalization layers, Leaky ReLu layers and dropout layers. The compressed representation is then passed to the subsequent decoder block, which consist of fully connected layers, normalization layers and Leaky ReLu layers. The Decoder outputs the reconstructed six seconds for the three channels. The details of the architecture are shown in Figure \ref{fig:det_arch}. With the reconstructed output an anomaly score is calculated by comparing it to the ground truth. A one-dimensional anomaly score is obtained by first calculating the \(L_2\) norm of each time point for the reconstructed output \(P = (p_1, p_2, \ldots, p_n)\) based on the three component values of each \(p_i = (x_1, x_2, x_3)\). This can be expressed mathematically as:

\begin{equation}
\| p_i \|_2 = \sqrt{x_1^2 + x_2^2 + x_3^2}
\end{equation}

The same calculation is performed for the ground truth values \(G = (g_1, g_2, \ldots, g_n)\), where each ground truth value \(g_i\) is also a three-dimensional vector \(g_i = (y_1, y_2, y_3)\):

\begin{equation}
\| g_i \|_2 = \sqrt{y_1^2 + y_2^2 + y_3^2}
\end{equation}

Finally, the resulting one-dimensional anomaly score time series is obtained by subtracting the computed \(L_2\) norms of the reconstructed outputs from those of the ground truth:

\begin{equation}
A = \| G \|_2 - \| P \|_2
\end{equation}
This anomaly time series will also be called anomaly score in subsequent sections.
In order to find the threshold for the anomaly score of the model the Gaussian background noise is calculated.

\subsection{Gaussian Background noise calculation}
 After calculating the anomaly score using the reconstructed output and ground truth, the next step is to find an appropriate threshold to differentiate between anomalous and standard values. For this, the method of finding the Background Gaussian Signal (BGS) is used. The following method as been proposed by Cuvier et al. \cite{bgs_paper} for the inspection of seismometer signal for signs of station degradation.
Given the assumption that continuous seismic signals follow a Gaussian distribution \cite{bgs_paper}. This distribution is often formulated as $X \sim \mathcal{N}(\sigma_0,\mu_0)$, $\sigma_0$ being the standard deviation and $\mu_0$ the mean. A form of characterizing the function is through the Cumulative Distribution Function (CDF). For a random variable X the CDF $\phi$ is defined as the probability that X takes a value less than or equal to a given real x. Another option for defining the standard Gaussian distribution is to use the inverse of the CDF, the quantile function, also called the Probit function, denoted as $\phi^{-1}$. To define non-standard Gaussian distributions with the Probit function, a translation is necessary of the mean $\mu_0$ and standard deviation $\sigma_0$.\cite{bgs_paper}
$\hat{\phi}^{-1} = \mu_0 + \sigma_0*\phi^{-1}$  

A practical approximation of the Probit function can be obtained through sorting the data points according to increasing values. Anomalies are visible as they stand out from the non-Gaussian noise data points. They gather at the start and end of the sorted set. In order to choose the thresholds $[Q_a, Q_b]$ for when to cut off the anomalies, the Probit function of each quantile interval in the sample space is compared with the modified Probit function. In the end the interval with the lowest difference between the two will be chosen. It will differ between deviations and Gaussian noise. As the goal is to work with continuous time data and to achieve fast detection this threshold mechanism is applied on a subset, with the length as well as the frequency of calculation being adjustable variables. 

\section{Data}\label{sec3}

\subsection{Training data}
The training data consists of noise recordings from 9 different Chilean stations. They comprise around 126 hours of data, with a sampling rate of 1 Hz, from the NSF GAGE Facility archives datasets \cite{UNAVCO_GPS}, with each station having around 14 hours of data, without the presence of ground displacements caused by an earthquake, that is, only noise. The used stations are: ANTC,\, CONS,\, FUTF,\, IQQE,\, JUNT,\, OVLL,\, PEDR,\, PTRO and QLLN. 

\begin{figure}[h]
\centering
\includegraphics[scale=0.5]{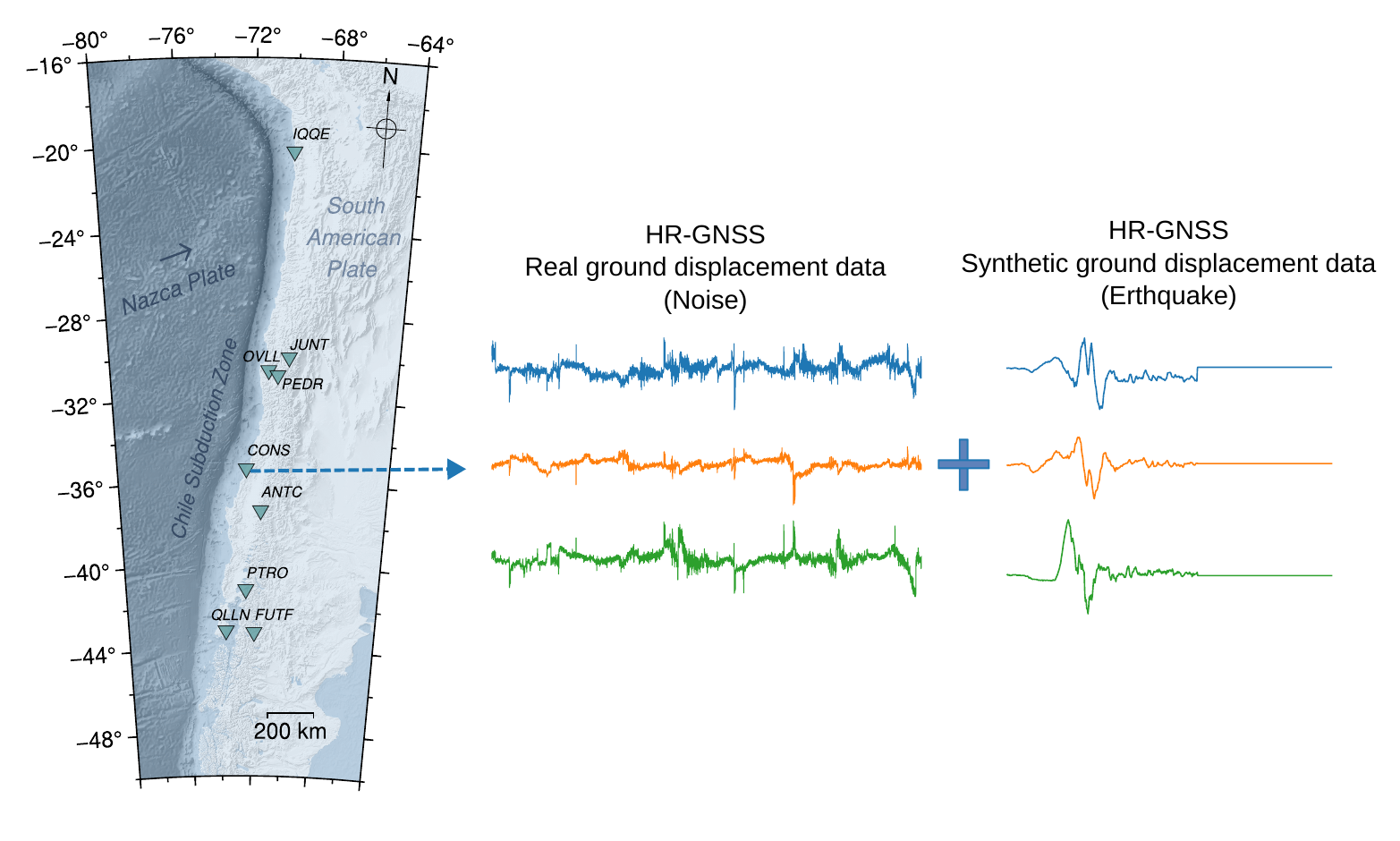}
\caption{Chilean stations that provide real, three-channel noise traces for training. From the same stations different noise traces are combined with synthetic events for validation and testing purposes.}
\label{fig:data_map}
\end{figure}
\subsection{Testing data}

\subsubsection{Synthetic data}
As HR-GNSS recordings of higher magnitude are scarce, artificially produced data is used for validation and testing purposes. These synthetic earthquakes consist of traces generated from the same 9 Chilean stations and have a length of 512 seconds \cite{bib61}. Since this synthetic data is free of noise it needs to be added by combining it with the seismic signals. The combined traces are 50000 seconds in length. Furthermore the data is labeled by hand to mark the start and end of seismic activity. Of these 9 stations with synthetic earthquakes eight are used as a first test set and one station is used for validation. Figure \ref{fig:data_map} shows an overview of the nine Chilean stations.

\subsubsection{Real data}
\subsubsection*{Data Processing by Multi-Constellation Satellite Positioning Observations}\label{sec4}

\subsubsection*{Observation Analysis-Diagnosis}
To ensure the quality of positional data, this procedure involves the detection of potential errors (noise). The methodology focuses on the identification and evaluation of satellite observation signals received at each measurement station. The acceptability of observations is determined systematically identifying and rejecting data that fall outside the specified response index. This process ensures that the resulting positional measurements achieve the required precision, with an accuracy on the order of 0.01 meters.

The analysis of satellite observation quality was conducted using the academic software GLab \cite{bib_glab}. This tool facilitates data quality verification by addressing various preprocessing challenges, such as the translation and editing of GNSS data.
Several parameters related to GNSS reception signals were analyzed, including multipath effects (MP), signal-to-noise ratio (SNR) \cite{bib_Bilich}, reception sensitivity in relation to the elevation angle, geometric factors such as GDOP, HDOP, VGDOP, and PGDOP, cycle slips, and potential atmospheric-induced errors.

\subsubsection*{Solution}
The solution for each epoch (HR-GNSS observation sampling) was obtained from GNSS phase observable measurements using the kinematic positioning mode. To validate the accuracy of the solutions, two programs were utilized: RTKLIB \cite{bib_Takasu} for Precise Point Positioning (PPP) \cite{bib_Zumberge}, and TRACK \cite{bib_Herring} for relative differentiation positioning (DP), using precise IGS final orbits and clock correction differences \cite{bib_Szolucha}.
The positions solution process incorporated ocean loading models from the finite element solutions (FES) tidal atlases \cite{bib_FES}. To mitigate atmospheric delay effects, the VMF1 troposphere mapping function was applied \cite{bib_VMF1}, while ionospheric noise was minimized using the Melbourne-Wübbena linear combination \cite{bib_Melbourne}, effectively reducing noise in the geodetic positions.

\subsubsection*{Adjustment}
The final position adjustment was performed using a mathematical-statistical procedure to compute residuals and their least squares for each position, resulting in coordinates within a defined reference framework. This method was employed to establish geodetic points.
The positions obtained using the PPP method were calculated for unfiltered data \cite{bib_Yigit}. Both PPP and DP solutions were used to generate time series of positions in geocentric Cartesian coordinates (X, Y, Z) within the ITRF2008 reference frame \cite{bib_Altamimi}, enabling the determination of displacements in each direction.
The PPP solutions \cite{bib_Teunissen} had a linear conformation, likely influenced by unmodeled characteristics of the observable signals. In contrast, the DP solutions minimized errors through the double-difference process.
For the generation of time series, 24-hour observation windows were utilized to extract displacements, calculated at one-second intervals.
The results from PPP and DP solutions for short-period observation windows were found to be similar. This outcome indicates that errors, primarily caused by atmospheric delays during satellite signal propagation, were minimal.

\section{Model Testing}\label{subsec4}

\subsection{Concepci\'on Earthquake}
\begin{figure}[h!]
\centering
\includegraphics[scale=0.8]{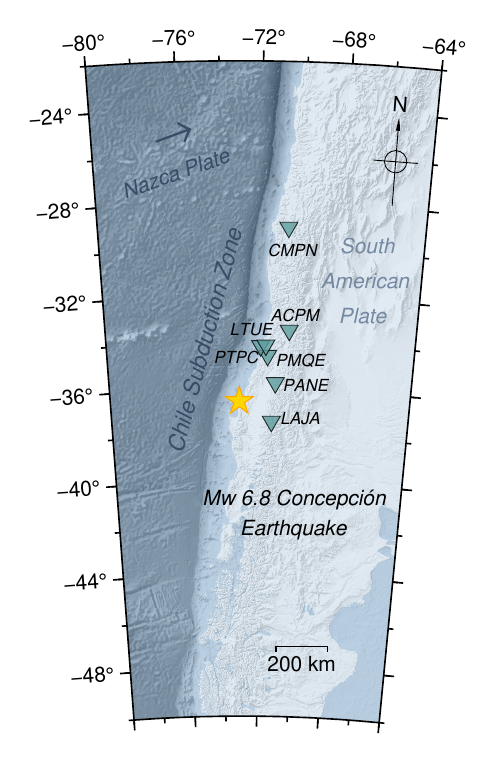}
\caption{Concepci\'on Earthquake epicenter and the six GNSS stations used for detection.}
\label{fig:concepcion_map}
\end{figure}

Real GNSS recordings of the earthquake near Concepci\'on, Chile are used for testing purposes of a real application scenario. The earthquake occurred on 11th February 2011 with a magnitude of 6.8. The traces encompass 24 hour long GNSS measurements of the earthquake, for six different stations from the Chilean National Seismological Network (CSN), which are available in the CDDIS data center \cite{CDDIS_data}. Figure \ref{fig:concepcion_map} shows a map of the stations and the event, Table \ref{tab:station_distances} has the accompanying information about station distance.

\begin{figure}[h!]
\centering
\includegraphics[scale=0.5]{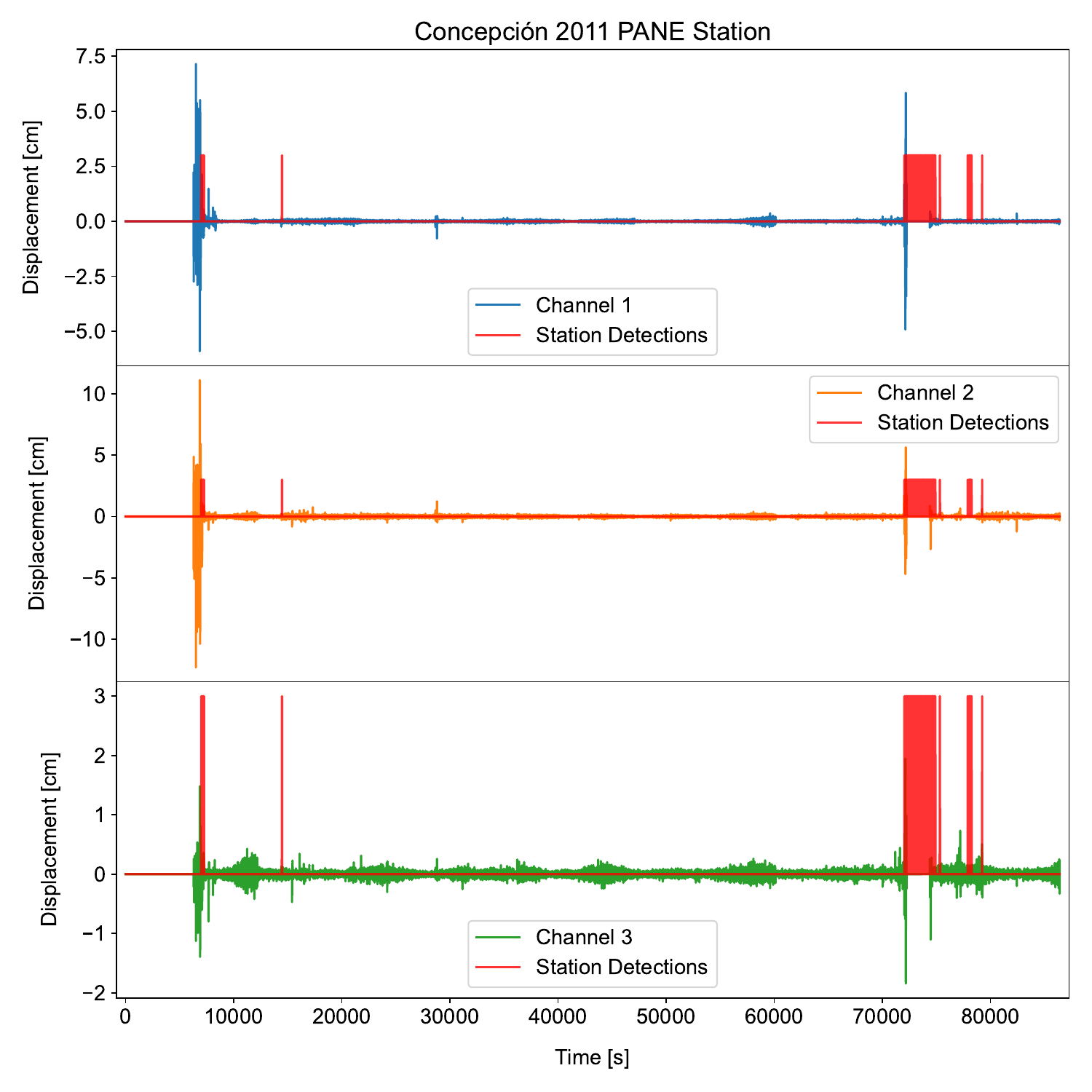}
\caption{24 hour trace of the PANE station. Multiple anomaly detections of the model are in red. The earthquake occurs at second 72000.}
\label{fig:pane_whole_w_start}
\end{figure}

\subsection{Overview}
The recording stations of the Concepci\'on Earthquake vary severely in their distance to the earthquake's epicenter, the nearest stations being PANE and LAJA. Choosing a threshold of approximately less than 3.5\textdegree guarantees that the event is still visible on the recording \cite{bib_GMM}. Stations further away can lead to less or even no recognizable displacement recordings. The 24 hour long time series allow a good assessment of the capabilities and limitations of the model. It also highlights some inherent challenges that the data source holds. Figure \ref{fig:pane_whole_w_start} shows an overview of the three channels of the station PANE. The detection of the model is shown in red, values above zero indicate detected anomalous timesteps. The first 7000 seconds are composed of missing data as well as data with calibration errors. Prior to the earthquake at the 72000 second there are some disturbances resembling earthquakes. 

\begin{table}[ht]
\centering
\begin{tabular}{|c|c|}
\hline
\textbf{Station} & \textbf{Distance [km]} \\
\hline
ACPM & 414.11 \\
LAJA & 181.75 \\
LTUE & 296.47 \\
PANE & 193.11 \\
PMQE & 256.74 \\
PTPC & 283.99 \\ 
\hline
\end{tabular}
\caption{Distances of the stations from the epicenter of the Concepci\'on earthquake.}
\label{tab:station_distances}
\end{table}

\subsection{Earthquake Detection}
\begin{figure}[ht!]
\centering
\includegraphics[scale=0.5]{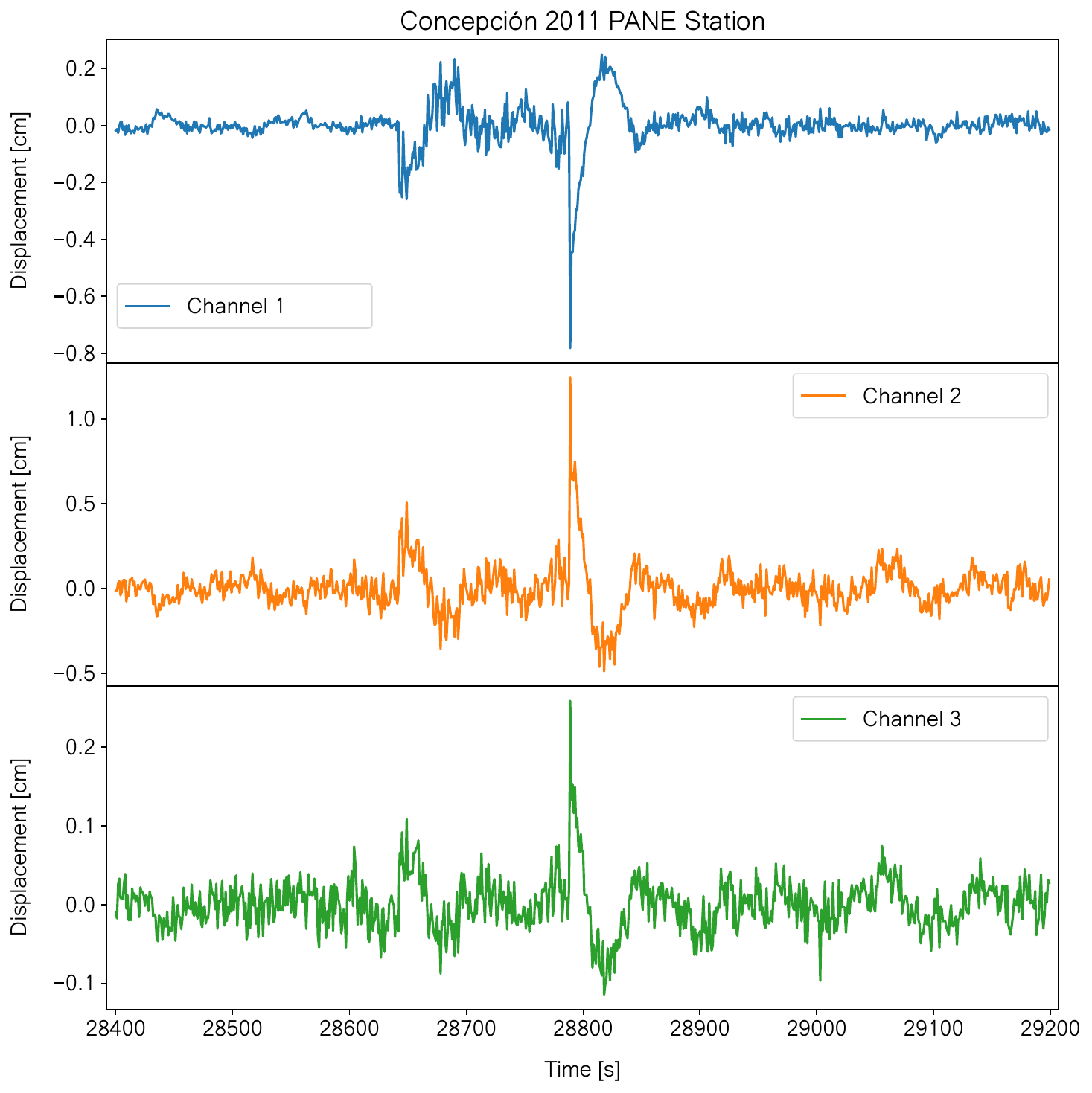}
\caption{Local disturbance at second 28800 recorded by the Chilean PANE station. The network does not detect an anomaly as not enough stations register the disturbance.}
\label{fig:local_disturb}
\end{figure}

Prior to the earthquake there are several potential false positives before 20000 seconds. Zooming in on one example, as shown in Figure \ref{fig:local_disturb}, reveals them as local disturbances with short duration, lasting only a few seconds.
It is important to note that the starting characteristics of this disturbance, such as the amplitude, mimic well a potential beginning of an earthquake. 

\begin{figure}[ht!]
\centering
\includegraphics[scale=0.5]{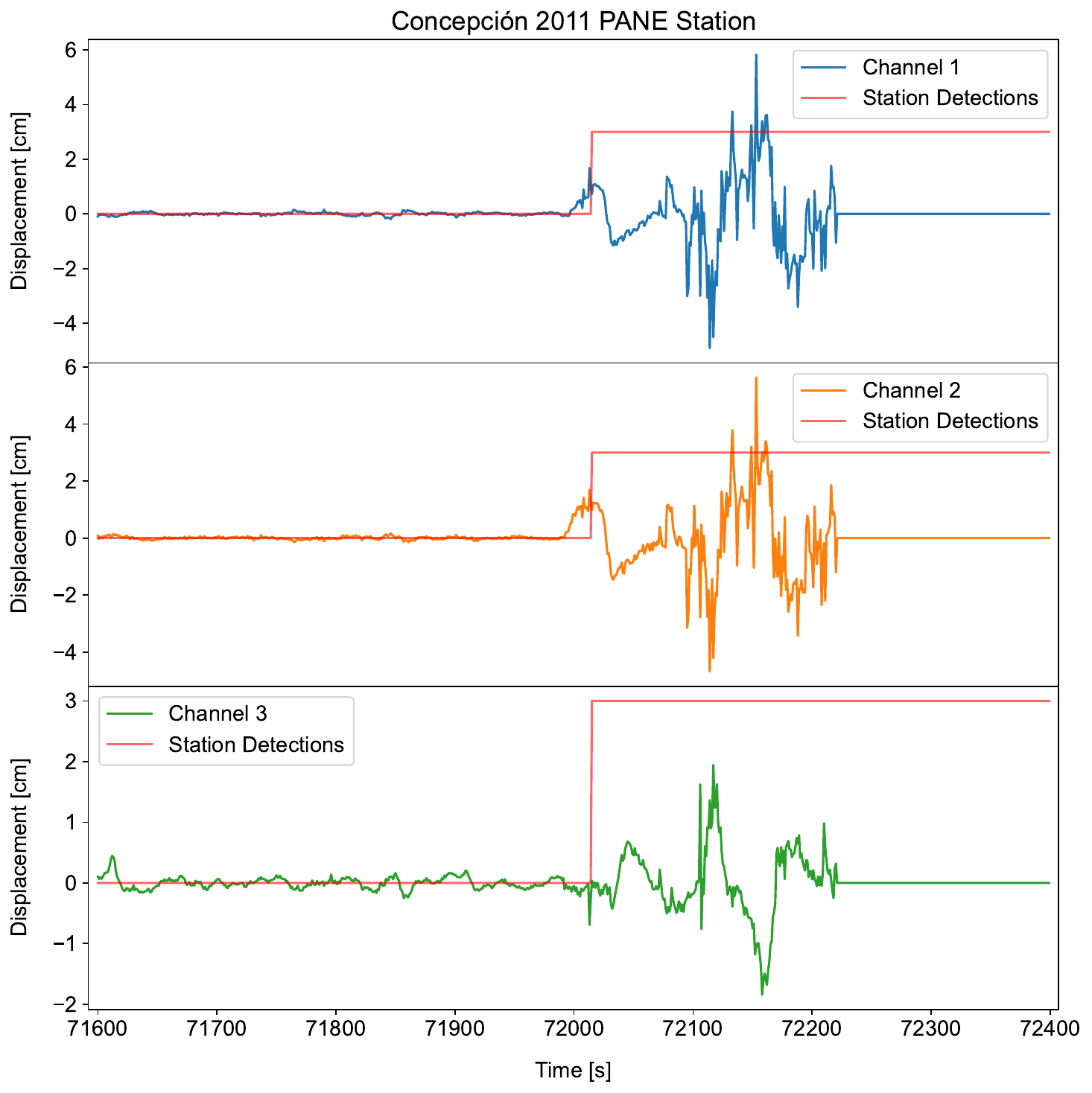}
\caption{Concepci\'on Earthquake recorded by the Chilean PANE station and DetEq's detection.}
\label{fig:pane_zoom_eqk1}
\end{figure}

In Figure \ref{fig:pane_zoom_eqk1} the beginning of the earthquake is visible, at second 72000. The detection from the model happens at second 72015 leaving a latency of 15 seconds.

 \subsection{Event Detection using Multiple Station}
Besides PANE, the rest of the five stations have the same characteristics regarding the starting calibration noise, some false detections and the correct detection of the earthquake with a varying latency. Special interest lies in the stations of LAJA and PANE as they are the closest. They would ideally be the first to record the seismic event and subsequently lead to the earliest detection. Furthermore, the seismic event should be more distinguishable in these traces. This is because the amplitude of the seismic waves diminish with increasing travel distance, changing the first waves to be indistinguishable from noise and creating a softer transition at the start of the earthquake. This complicates the recognition of the event.

\begin{figure}[ht!]%
    \centering
    \subfloat[\centering Counted false detection prior to the earthquake]{{\includegraphics[width=6cm]{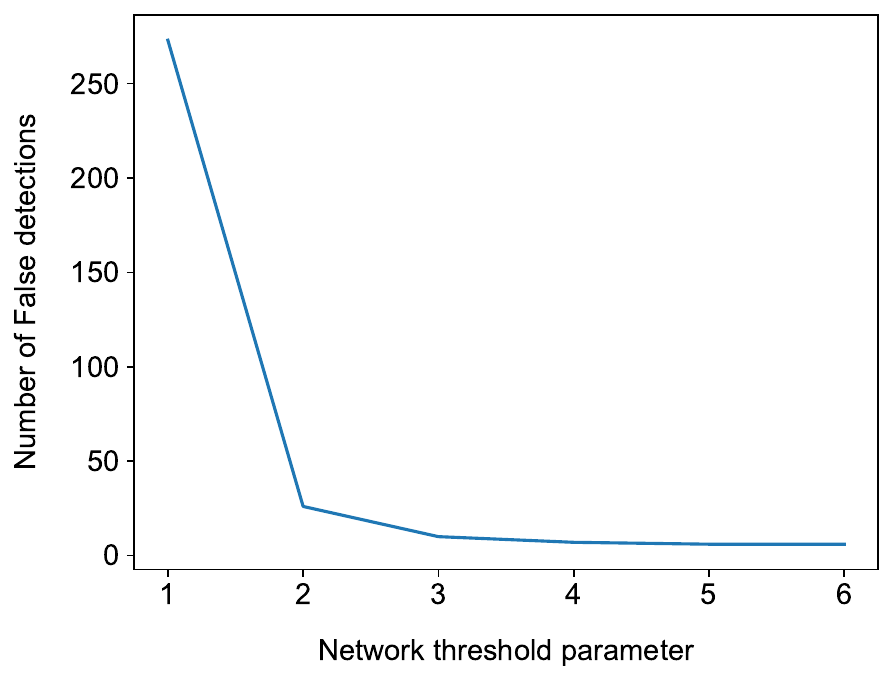} }}%
    \qquad
    \subfloat[\centering Seconds of late detection.]{{\includegraphics[width=6cm]{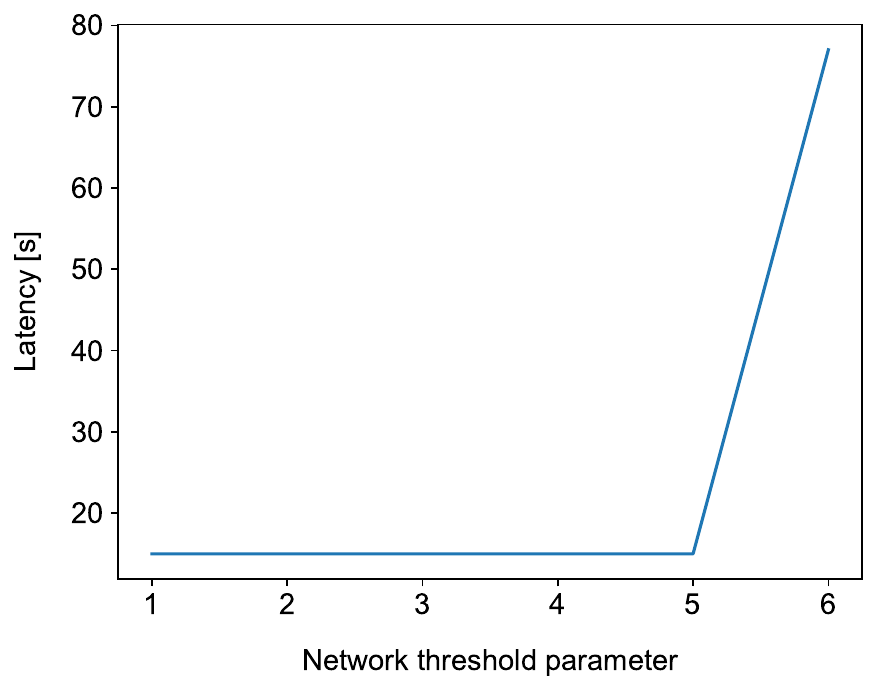} }}%
    \caption{Network detection threshold parameter search for the seven station network of the Concepci\'on earthquake}%
    \label{fig:analisis_both}%
\end{figure}

 One configuration parameter of the pipeline is the minimal required length of an detection before it is considered a real anomaly. This depends heavily on the use case. Another parameter allows to sort out detections that lack accompanying later detections. Note that through this method the false detections can be minimized, however this comes at the cost of detection speed. Several seconds need to pass in order to analyze the properties of duration and detection count. These methods are still useful for other use cases that do not involve detection in real time. 
 For ensuring fast, real time detection the deployment of the pipeline is expanded to a network of stations. Having the additional resources of a good station network allows the detection pipeline to run on multiple stations at the same time and cross reference each detection. Local disturbances are not visible on other stations while the seismic event is sure to arrive within seconds to all nearby stations. For this we propose an additional configurable parameter, the length of allowed latency between stations.
 This is necessary as the earthquake signal will arrive at different times at each station, depending on their distance to the epicenter. The proposed allowed latency parameter defines if these detections belong to the same event or not. The parameter could depend on different factors such as the distance between stations and geological properties of the region. In the experiment with the Concepci\'on earthquake a fixed window of 15 seconds before and after a detection is used. The network threshold parameter decides how many stations in the network need to detect an anomaly in order to be regarded as an earthquake. The Figure \ref{fig:analisis_both} (a and b) show that choosing five as a parameter for the six station network lead to the least false detections prior to the earthquake while retaining fast detection time. 

\section{Conclusion and Future Work}\label{sec6}
The DL model successfully detected seismic events across the displacement observations from multiple GNSS stations. We demonstrated the performance of the DetEq model using the 2011 Mw 6.8 Concepci\'on  earthquake as a case study. Additionally, the model has been tested on other earthquakes, including events in Chile and Mexico, such as the Mw 7.4 Oaxaca earthquake in 2020 (see Supplementary Information). These results highlight the versatility and reliability of the DetEq model for detecting large seismic events across diverse tectonic settings.

The detection pipeline includes configurable parameters to optimize for different use cases. The minimal detection length and detection-count thresholds effectively reduce false positives but introduce delays in detection speed. These trade-offs make the approach suitable for post-event analysis or slower detection scenarios, though they are less ideal for real-time applications.

To enhance real-time detection capabilities, deploying the pipeline across a particular network of stations is recommended. This setup allows for cross-referencing detections, minimizing the impact of local disturbances at individual stations. The additional configurable parameter, allowed latency between stations, is critical to account for the variable arrival times of seismic signals at different stations. This parameter can be tailored based on station distances, local velocity properties, and other geophysical factors.

The results underscore the potential of GNSS-based DL models in earthquake detection, particularly when integrated into a robust station network. Further research on improving real-time detection capabilities is essential to minimize false detections while maintaining rapid detection times for practical implementation in early warning systems.

The results of this study lay the groundwork for creating a  GNSS data based earthquake monitoring pipeline which will be incorporated into the SAIPy package \cite{bib_saipy}. This future pipeline will not only include the proposed earthquake detection model (DetEq) but will also integrate DL-based magnitude estimation models \cite{bib_gnss_egu}. Thus, SAIPy will offer an open-source, modular, and extensible platform for GNSS-based seismological research.

\subsection*{Code availability}
All codes in this work are available at https://github.com/srivastavaresearchgroup/SAIPy. The DetEq model is implemented in PyTorch \cite{pytorch}.

\subsection*{Acknowledgments}
 This research was supported by the Federal Ministry of Education and Research of Germany (BMBF), grant SAI 01IS20059. Modeling and data processing were performed at the Frankfurt Institute for Advanced Studies, with a GPU cluster funded by BMBF for the project Seismologie und Artifizielle Intelligenz (SAI). The training data used in this study have been provided by the GAGE Facility, operated by UNAVCO, Inc., with support from the National Science Foundation, the National Aeronautics and Space Administration, and the U.S. Geological Survey under NSF Cooperative Agreement EAR-1724794.

\subsection*{Author Contributions}
Conceptualization: Javier Quintero-Arenas, Claudia Quinteros-Cartaya, Nishtha Srivastava; Data processing and preparation: Javier Quintero-Arenas, Claudia Quinteros-Cartaya, Andrea Padilla-Lafarga, Carlos Moraila; Methodology: Javier Quintero-Arenas, Claudia Quinteros-Cartaya, Johannes Faber, Jonas Köhler, Nishtha Srivastava; Formal analysis and investigation: Javier Quintero-Arenas; Validation: Javier Quintero-Arenas, Claudia Quinteros-Cartaya, Andrea Padilla-Lafarga, Carlos Moraila; Writing ‐ original draft preparation: Javier Quintero-Arenas, Claudia Quinteros-Cartaya,  Andrea Padilla-Lafarga, Carlos Moraila; Writing ‐ review and editing: Javier Quintero-Arenas, Claudia Quinteros-Cartaya, Andrea Padilla-Lafarga, Carlos Moraila, Johannes Faber, Jonas Köhler, Nishtha Srivastava; Funding acquisition: Nishtha Srivastava; Supervision: Nishtha Srivastava

\bibliography{references}

\end{document}